\PassOptionsToPackage{numbers}{natbib}
\documentclass[12pt]{article}

\usepackage[preprint]{neurips_2025}
\usepackage{array}
\usepackage{multirow}
\usepackage{booktabs}
\usepackage{amssymb}
\usepackage{pifont}

\usepackage[protrusion=false, expansion=false]{microtype}

\usepackage{hyperref}
\usepackage{url}
\usepackage{amsfonts}
\usepackage{nicefrac}
\usepackage{xcolor}
\definecolor{clBlue}{HTML}{4A7FB5}
\definecolor{clBlueBg}{HTML}{E8F0FA}
\definecolor{t1Orange}{HTML}{E8772E}
\definecolor{t1OrangeBg}{HTML}{FFF3E8}
\definecolor{stage1Color}{HTML}{5B9BD5}
\definecolor{stage2Color}{HTML}{70AD47}
\definecolor{stage3Color}{HTML}{ED7D31}
\definecolor{arrowGray}{HTML}{888888}
\usepackage{parskip}
\usepackage{setspace}
\usepackage{titlesec}

\titlespacing*{\section}{0pt}{0.3em plus 0.1em minus 0.1em}{0.3em}
\titlespacing*{\subsection}{0pt}{0.3em plus 0.1em minus 0.1em}{0.3em}
\titlespacing*{\subsubsection}{0pt}{0.8em}{0.5em}

\titleformat{\section}
  {\normalfont\fontsize{14pt}{17pt}\bfseries}
  {\thesection}{1em}{}

\titleformat{\subsection}
  {\normalfont\fontsize{12pt}{15pt}\bfseries}
  {\thesubsection}{1em}{}

\titleformat{\subsubsection}
  {\normalfont\fontsize{11pt}{14pt}\bfseries}
  {\thesubsubsection}{1em}{}

\title{CRE-T1 Preview Technical Report: Beyond Contrastive Learning for Reasoning-Intensive Retrieval}
\usepackage{graphicx}
\usepackage{tikz}
\usetikzlibrary{arrows.meta, positioning, calc, fit, backgrounds, decorations.pathreplacing, shapes.geometric}
\usepackage{fancyvrb}
\usepackage[most]{tcolorbox}
\usepackage{listings}

\newtcolorbox{promptblock}{colback=gray!8,colframe=gray!60,fontupper=\ttfamily\footnotesize,breakable}

\author{%
  \textbf{Guangzhi Wang} \quad \textbf{Yinghao Jiao} \quad \textbf{Zhi Liu}\\[0.5em]
  CareerInternational Research Team \\[0.5em]
}

\begin{document}
\maketitle

\begin{abstract}
The central challenge of reasoning-intensive retrieval lies in identifying implicit reasoning relationships between queries and documents, rather than superficial semantic or lexical similarity. The contrastive learning paradigm is fundamentally a static representation consolidation technique: during training, it encodes hierarchical relevance concepts into fixed geometric structures in the vector space, and at inference time it cannot dynamically adjust relevance judgments according to the specific reasoning demands of each query. Consequently, performance degrades noticeably when vocabulary mismatch exists between queries and documents or when implicit reasoning is required to establish relevance. This paper proposes Thought 1 (T1), a generative retrieval model that shifts relevance modeling from static alignment to dynamic reasoning. On the query side, T1 dynamically generates intermediate reasoning trajectories for each query to bridge implicit reasoning relationships and uses \texttt{<emb\_token>} as a semantic aggregation point for the reasoning output. On the document side, it employs an instruction + text + \texttt{<emb\_token>} encoding format to support high-throughput indexing. To internalize dynamic reasoning capabilities into vector representations, we adopt a three-stage training curriculum and introduce GRPO in the third stage, enabling the model to learn optimal derivation strategies for different queries through trial-and-error reinforcement learning. On the BRIGHT benchmark, T1-4B exhibits strong performance under the original query setting, outperforming larger models trained with contrastive learning overall, and achieving performance comparable to multi-stage retrieval pipelines. The results demonstrate that replacing static representation alignment with dynamic reasoning generation can effectively improve reasoning-intensive retrieval performance.
\end{abstract}

\section{Introduction}

Since the popularization of bi-encoder architectures such as Sentence-BERT \cite{reimers2019sentence}, contrastive learning \cite{gao2021simcse} has become the standard paradigm for training dense vector models. However, in reasoning-intensive retrieval tasks, merely aligning ``similarity'' is often insufficient to capture ``relevance.'' The key challenge in such tasks is that the evidence supporting an answer frequently exhibits vocabulary mismatch with the query, requiring implicit causal, conditional, or multi-hop reasoning relationships to establish connections. Contrastive learning typically uses the representational distance between positive and negative samples as the supervisory signal, which tends to reward lexical matching and shallow semantic consistency while exhibiting noticeable performance degradation in scenarios requiring implicit reasoning. Examining this limitation from a deeper perspective, it is not accidental. Contrastive learning is fundamentally a static, hierarchical representation consolidation process: during training, it encodes the hierarchical conceptual judgment of ``relevant/irrelevant'' into the geometric structure of the vector space through pre-constructed positive-negative sample pairs, completing a one-way mapping from concepts to latent variables. Once training converges, this mapping is frozen as a fixed prior in the model parameters---a pre-training variable---and at inference time the model can only retrieve within this fixed representation space, unable to dynamically adjust its relevance judgment strategy according to the specific reasoning demands of a new query. This problem is particularly prominent in reasoning-intensive benchmarks such as BRIGHT \cite{su2024bright}: a large number of queries require the model to identify implicit reasoning relationships rather than simple keyword or shallow semantic matches, yet the static representation space consolidated by contrastive learning struggles to cover these implicit reasoning relationships that require per-query derivation.

To address this challenge, we propose the retrieval model Thought 1 (T1). Its core idea is to replace static alignment with dynamic reasoning: without significantly increasing retrieval cost, the model first generates a reasoning trajectory for each query and then compresses the reasoning result into a retrievable vector, so that relevance judgments are grounded in explicit derivation rather than solely relying on static representations consolidated during training. To motivate this design choice, we draw on our perspective on agent cognition: the essence of human problem-solving under limited cognitive resources lies in temporalized sequential reasoning \cite{liu2026timescaling}. Therefore, ``scaling computation along the temporal dimension'' is a key lever for improving capability. Based on this argument, related approaches can be naturally divided into two complementary lines. The first is \textit{test-time scaling}, which invests more computation at inference time in exchange for accuracy; our prior work Refine Thought (RT) \cite{wang2025refine} belongs to this category, iteratively refining semantic representations in hidden space through multiple forward passes. The second is \textit{train-time scaling}, which consolidates dynamic reasoning capabilities into model parameters during training through stronger supervision or reinforcement learning; for example, DeepSeek-R1 \cite{deepseekr1} enhances logical capabilities by optimizing reasoning trajectories. T1 adopts reinforcement learning at training time as its core mechanism, enabling the model to learn the ability to dynamically generate reasoning paths conditioned on each query, thereby transforming relevance modeling in reasoning-intensive retrieval from static representation alignment into a dynamic reasoning generation process.

Experiments demonstrate that T1 exhibits significant advantages in reasoning-intensive retrieval scenarios such as BRIGHT. The key lies in fusing the query understanding and text representation stages of the retrieval pipeline: after observing the query constraints, the model first reasons, then compresses both the reasoning and query semantics into a query vector, which is compared with document vectors via similarity computation to complete retrieval. To achieve this, T1 adopts an asymmetric architecture: the query side generates a limited-length reasoning sequence and aggregates global semantic information through \texttt{<emb\_token>}, while the document side extracts the corresponding token's representation via instructions to ensure large-scale index construction efficiency. We internalize the aforementioned reasoning pattern into model parameters through reinforcement learning during training, thereby improving reasoning-intensive relevance modeling without significantly increasing online overhead. On the BRIGHT evaluation benchmark, T1 at the 4B scale using only original queries outperforms the combination of ``retrieval model + query rewriting,'' significantly surpasses contrastive baselines and commercial embedding models, and matches the state-of-the-art ``retrieval model + reranking'' combination pipeline.

\section{Related Work}

We define reasoning-intensive retrieval as tasks where the relevance between a query and its relevant documents depends on implicit reasoning chains (semantic, causal, or logical) rather than mere keyword overlap or shallow semantic similarity. Early and mainstream dense retrieval benchmarks (e.g., BEIR \cite{thakur2021beir}, MTEB \cite{muennighoff-etal-2023-mteb}) primarily focus on Information Seeking queries, where answers typically exist explicitly within text passages, or queries and documents exhibit high semantic match in vector space. The recent BRIGHT benchmark \cite{su2024bright} further systematizes retrieval complexity into three levels: Level 1 relies on lexical matching, Level 2 relies on semantic matching, and Level 3 requires the model to locate evidence through implicit reasoning relationships. Compared with the first two levels, Level 3 tasks better reveal differences in models' reasoning-based relevance modeling capabilities. For instance, in BRIGHT, some models that lead on MTEB (e.g., BGE \cite{bge2023}) exhibit significant drops in nDCG@10, highlighting the gap between semantic matching evaluation and reasoning-intensive relevance.

Existing work on reasoning-intensive retrieval can be broadly categorized into three technical lines based on the source of training signals and where reasoning augmentation occurs. The first line addresses reasoning signal sparsity from the perspective of \textit{data and supervision}: it enhances the model's representation learning on complex contexts through synthetic reasoning data or structured supervision, with representative works including ReasonIR \cite{shao2025reasonir}, RaDeR \cite{das2025rader}, and ReasonEmbed \cite{chen2025reasonembed}. The second line targets \textit{retrieval pipeline decomposition}, introducing reasoning augmentation at different stages: on the query understanding end, generative intent decomposition improves complex question parsing, e.g., DeepRetrieval \cite{jiang2025deepretrieval}, Reinforced-IR \cite{li2025reinforced}, TongSearch-QR \cite{qin2025tongsearch}; on the ranking end, reasoning chain generation is combined with reranking to improve global consistency, e.g., Rank1 \cite{weller2025rank1}, Rank-K \cite{yang2025rankk}, ReasonRank \cite{liu2025reasonrank}. The third line models retrieval as a \textit{dynamic decision process}, using increased test-time compute and reinforcement learning training strategies to decide when to retrieve and how to reformulate intent, with Search-R1 \cite{jin2025searchr1} as a representative work.

\section{Method}

\subsection{Problem Definition}

In reasoning-intensive retrieval tasks, given a query $q$ and a document collection $\mathcal{D}$, the goal is to recall relevant documents $d \in \mathcal{D}$. Dense retrieval typically ranks candidates by the similarity between query and document embeddings, with the scoring function defined as $s(q,d)=\mathrm{sim}(\mathbf{e}_q,\mathbf{e}_d)$ (e.g., inner product or cosine similarity). T1 fuses the query understanding and text representation stages of the retrieval pipeline end-to-end within a single model and adopts an asymmetric architecture to encode queries and documents separately. We introduce a special token \texttt{<emb\_token>} into the vocabulary as a semantic aggregation point. On the query side, the model generates a reasoning sequence $\mathbf{r}_{1:L}$ via limited-step autoregressive decoding and learns during training to automatically output \texttt{<emb\_token>} upon completing the reasoning; we take the hidden state at this token position as the query vector:
\[
  \mathbf{e}_q = \mathbf{h}_\theta\!\left([\;q,\ \mathbf{r}_{1:L},\ \texttt{<emb\_token>}\;]\right)_{\texttt{<emb\_token>}}
\]
On the document side, the model does not generate a reasoning sequence. Instead, it concatenates a document-side instruction $\mathcal{I}_d$ with the document content $d$, appends \texttt{<emb\_token>} at the end, and performs a single encoding-style forward pass to extract the corresponding hidden state as the document vector:
\[
  \mathbf{e}_d = \mathbf{h}_\theta\!\left([\;\mathcal{I}_d,\ d,\ \texttt{<emb\_token>}\;]\right)_{\texttt{<emb\_token>}}
\]
This design ensures indexing throughput while enabling the query-side ``reason-then-represent'' process to explicitly participate in vector generation.

The above formulation reveals the fundamental difference between contrastive learning and T1 in relevance modeling, which can be distilled as the opposition between static and dynamic paradigms. Under the standard contrastive learning framework, the query representation $\mathbf{e}_q = f_\theta(q)$ is a deterministic mapping of the input $q$, and the training objective optimizes the average similarity distance over pre-constructed positive-negative pairs $\{(q_i, d_i^+, d_i^-)\}$, encoding hierarchical relevance concepts (``relevant / partially relevant / irrelevant'') into fixed geometric structures in the vector space. This process is essentially a one-way consolidation from concepts to latent variables: the decision boundary for relevance is determined within the statistical distribution of the training set and then frozen as a parametric prior. At inference time, the model can only perform nearest-neighbor search on this fixed manifold and cannot generate new association paths tailored to the reasoning demands of a specific query. When vocabulary mismatch exists between queries and documents or multi-hop derivation is required, this static representation space struggles to cover implicit reasoning relationships outside the training distribution, degenerating into a strategy that favors lexical or shallow semantic matching.

The core design of T1 is precisely to shift relevance modeling from a static mapping to a dynamic reasoning process. In T1, the reasoning sequence $\mathbf{r}_{1:L}$ is a query-conditioned autoregressive generation: each query triggers an independent reasoning path, and the model dynamically constructs a bridging chain from the question to the evidence based on the semantic constraints and latent logical demands of the current query. GRPO further enables the model to learn to dynamically select the optimal derivation strategy for different queries through trial-and-error reinforcement learning, rather than passively aligning representations on fixed positive-negative pairs. This paradigm shift---from ``pre-encoding fixed relevance patterns'' to ``on-demand reasoning path generation''---ensures that the query representation is no longer a static function of the input but a dynamic product of the reasoning process, aggregating reasoning and semantic information into a retrievable vector representation through \texttt{<emb\_token>}.

\subsection{Training Details}\label{sec:training}

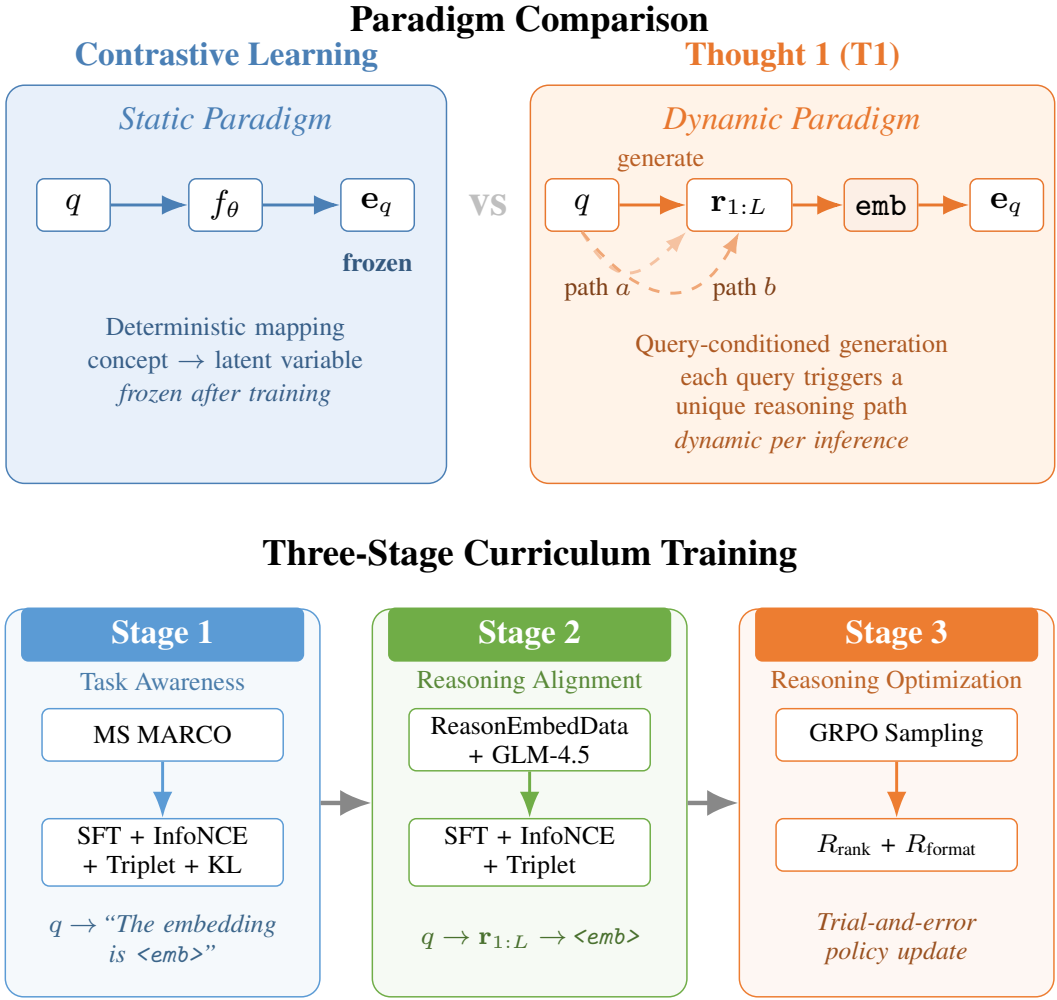
\begin{figure}[ht]
  \centering
  \resizebox{\linewidth}{!}{%
  \begin{tikzpicture}[
    every node/.style={font=\small},
    cbox/.style={draw, rounded corners=2pt, minimum height=0.5cm, minimum width=0.7cm,
                align=center, font=\footnotesize, inner sep=2pt},
    sbox/.style={draw, rounded corners=2pt, minimum height=0.5cm, minimum width=1.0cm,
                 align=center, font=\footnotesize, inner sep=2pt},
    myarrow/.style={-{Latex[length=2.5mm]}, thick, #1},
    myarrow/.default={arrowGray},
    bigarrow/.style={-{Latex[length=3mm, width=2.5mm]}, very thick, arrowGray},
    titlefont/.style={font=\bfseries\small},
  ]

  \node[titlefont, font=\bfseries] at (5, 0.6) {Paradigm Comparison};

  \node[draw=clBlue, fill=clBlueBg, rounded corners=5pt,
        minimum width=4.2cm, minimum height=3.8cm, anchor=north west,
        label={[clBlue, titlefont]above:Contrastive Learning}]
        (clBox) at (0, 0) {};
  \node[font=\footnotesize\itshape, clBlue] at (2.1, -0.35) {Static Paradigm};

  \node[cbox, fill=white, draw=clBlue] (cQ) at (0.65, -1.15) {$q$};
  \node[cbox, fill=white, draw=clBlue] (cF) at (2.1, -1.15) {$f_\theta$};
  \node[cbox, fill=white, draw=clBlue] (cE) at (3.55,-1.15) {$\mathbf{e}_q$};
  \draw[myarrow=clBlue] (cQ) -- (cF);
  \draw[myarrow=clBlue] (cF) -- (cE);

  \node[font=\scriptsize\bfseries, clBlue!70!black] at (3.55, -1.7) {frozen};

  \node[font=\scriptsize, text width=3.6cm, align=center, clBlue!75!black]
    at (2.1, -2.65)
    {Deterministic mapping\\[1pt]
     concept $\rightarrow$ latent variable\\[1pt]
     \textit{frozen after training}};

  \node[font=\large\bfseries, gray!50] at (4.6, -1.15) {vs};

  \node[draw=t1Orange, fill=t1OrangeBg, rounded corners=5pt,
        minimum width=5.0cm, minimum height=3.8cm, anchor=north west,
        label={[t1Orange, titlefont]above:Thought 1 (T1)}]
        (t1Box) at (5.0, 0) {};
  \node[font=\footnotesize\itshape, t1Orange] at (7.5, -0.35) {Dynamic Paradigm};

  \node[cbox, fill=white, draw=t1Orange]                       (tQ)   at (5.5,  -1.15) {$q$};
  \node[cbox, fill=white, draw=t1Orange, minimum width=1.0cm]  (tR)   at (7.0,  -1.15) {$\mathbf{r}_{1:L}$};
  \node[cbox, fill=t1Orange!12, draw=t1Orange]                 (tTok) at (8.35, -1.15) {\texttt{emb}};
  \node[cbox, fill=white, draw=t1Orange]                       (tE)   at (9.55, -1.15) {$\mathbf{e}_q$};
  \draw[myarrow=t1Orange] (tQ)   -- (tR);
  \draw[myarrow=t1Orange] (tR)   -- (tTok);
  \draw[myarrow=t1Orange] (tTok) -- (tE);

  \node[font=\scriptsize, t1Orange!80!black] at ($(tQ)!0.5!(tR)+(0,0.42)$) {generate};

  \draw[t1Orange!45, thick, dashed, -{Latex[length=2mm]}]
    (tQ.south) .. controls +(0.3,-0.5) and +(-0.5,-0.5) .. (tR.south west);
  \draw[t1Orange!65, thick, dashed, -{Latex[length=2mm]}]
    (tQ.south) .. controls +(0.6,-0.75) and +(-0.4,-0.75) .. (tR.south);
  \node[font=\scriptsize, t1Orange!55!black] at (5.65, -1.95) {path $a$};
  \node[font=\scriptsize, t1Orange!55!black] at (7.05,  -1.95) {path $b$};

  \node[font=\scriptsize, text width=4.4cm, align=center, t1Orange!75!black]
    at (7.5, -2.95)
    {Query-conditioned generation\\[1pt]
     each query triggers a unique reasoning path\\[1pt]
     \textit{dynamic per inference}};

  \node[titlefont, font=\bfseries] at (5, -4.5) {Three-Stage Curriculum Training};

  \node[draw=stage1Color, fill=stage1Color!6, rounded corners=5pt,
        minimum width=3.0cm, minimum height=3.7cm, anchor=north]
        (s1box) at (1.5, -5.0) {};
  \node[fill=stage1Color, text=white, rounded corners=2pt,
        font=\footnotesize\bfseries, minimum width=2.7cm]
        at (1.5, -5.25) {Stage 1};
  \node[font=\scriptsize, stage1Color!80!black] at (1.5, -5.7) {Task Awareness};
  \node[sbox, fill=white, draw=stage1Color, minimum width=2.3cm, font=\scriptsize,
        anchor=north]
    (d1) at (1.5, -5.95) {MS MARCO};
  \node[sbox, fill=white, draw=stage1Color, minimum width=2.3cm, font=\scriptsize,
        text width=1.9cm, align=center, anchor=north]
    (l1) at (1.5, -7.0) {SFT + InfoNCE\\+ Triplet + KL};
  \node[font=\scriptsize\itshape, stage1Color!65!black, text width=2.5cm, align=center]
    at (1.5, -8.15) {$q \!\to\!$ ``The embedding\\is \texttt{<emb>}''};
  \draw[myarrow=stage1Color] (d1) -- (l1);

  \node[draw=stage2Color, fill=stage2Color!6, rounded corners=5pt,
        minimum width=3.0cm, minimum height=3.7cm, anchor=north]
        (s2box) at (5.0, -5.0) {};
  \node[fill=stage2Color, text=white, rounded corners=2pt,
        font=\footnotesize\bfseries, minimum width=2.7cm]
        at (5.0, -5.25) {Stage 2};
  \node[font=\scriptsize, stage2Color!80!black] at (5.0, -5.7) {Reasoning Alignment};
  \node[sbox, fill=white, draw=stage2Color, minimum width=2.3cm, font=\scriptsize,
        text width=1.9cm, align=center, anchor=north]
    (d2) at (5.0, -5.95) {ReasonEmbedData\\+ GLM-4.5};
  \node[sbox, fill=white, draw=stage2Color, minimum width=2.3cm, font=\scriptsize,
        text width=1.9cm, align=center, anchor=north]
    (l2) at (5.0, -7.0) {SFT + InfoNCE\\+ Triplet};
  \node[font=\scriptsize\itshape, stage2Color!65!black, text width=2.5cm, align=center]
    at (5.0, -8.15) {$q \!\to\! \mathbf{r}_{1:L} \!\to\!$ \texttt{<emb>}};
  \draw[myarrow=stage2Color] (d2) -- (l2);

  \node[draw=stage3Color, fill=stage3Color!6, rounded corners=5pt,
        minimum width=3.0cm, minimum height=3.7cm, anchor=north]
        (s3box) at (8.5, -5.0) {};
  \node[fill=stage3Color, text=white, rounded corners=2pt,
        font=\footnotesize\bfseries, minimum width=2.7cm]
        at (8.5, -5.25) {Stage 3};
  \node[font=\scriptsize, stage3Color!80!black] at (8.5, -5.7) {Reasoning Optimization};
  \node[sbox, fill=white, draw=stage3Color, minimum width=2.3cm, font=\scriptsize,
        anchor=north]
    (d3) at (8.5, -5.95) {GRPO Sampling};
  \node[sbox, fill=white, draw=stage3Color, minimum width=2.3cm, font=\scriptsize,
        text width=1.9cm, align=center, anchor=north]
    (l3) at (8.5, -7.0) {$R_{\text{rank}}$ + $R_{\text{format}}$};
  \node[font=\scriptsize\itshape, stage3Color!65!black, text width=2.5cm, align=center]
    at (8.5, -8.15) {Trial-and-error\\policy update};
  \draw[myarrow=stage3Color] (d3) -- (l3);

  \draw[bigarrow] (s1box.east) -- (s2box.west);
  \draw[bigarrow] (s2box.east) -- (s3box.west);

  \end{tikzpicture}%
  }
  \caption{Overview of the T1 model. Top: paradigm comparison between contrastive learning (static) and T1 (dynamic); Bottom: three-stage curriculum training pipeline (see Section~\ref{sec:training} for details).}
  \label{fig:architecture}
\end{figure}

We use Qwen3-Instruct-4B \cite{yang2025qwen3} as the base model for T1 and add \texttt{<emb\_token>} to its vocabulary. Training follows a three-stage curriculum (Figure~\ref{fig:architecture}): Stage~1 performs cold-start initialization to establish task awareness and output formatting; Stage~2 conducts reasoning alignment through reasoning data, enabling the model to generate reasoning sequences before outputting \texttt{<emb\_token>}; Stage~3 further optimizes reasoning quality and preference consistency under a reinforcement learning framework.

\paragraph{Stage 1: Task Awareness \& Formatting}
Stage 1 uses the MS MARCO dataset to establish foundational retrieval capability and consolidates the output format as ``Query $\rightarrow$ reasoning preamble $\rightarrow$ \texttt{<emb\_token>}.'' The objective of Stage 1 is cold-start initialization and format consolidation, so the SFT weight is increased accordingly. Meanwhile, InfoNCE/Triplet losses are retained to establish retrieval discriminative signals. The Triplet term uses a larger coefficient due to its smaller gradient magnitude, and the KL term serves only as a weak regularizer to stabilize training. The total loss is:
\[
\mathcal{L}_{\text{stage1}}
=
0.8\,\mathcal{L}_{\text{sft}}
+ 1.0\,\mathcal{L}_{\text{nce}}
+ 15\,\mathcal{L}_{\text{tri}}
+ 0.02\,\mathcal{L}_{\text{kl}}
\]

\paragraph{Stage 2: Reasoning Alignment (SFT)}
Stage 2 introduces the synthetic reasoning data from ReasonEmbed \cite{chen2025reasonembed} and uses GLM-4.5 to regenerate higher-quality and more concise reasoning paths, performing CoT injection so that the model generates a concise reasoning sequence (i.e., $\mathbf{r}_{1:L}$ from the Problem Definition) before outputting \texttt{<emb\_token>}, thereby explicitly externalizing the latent derivation process and incorporating it into vector generation. Compared with Stage 1, this stage shifts from ``format consolidation + basic discrimination'' to ``reasoning alignment,'' so the SFT weight is increased to strengthen supervision over the reasoning text and output interface, while the Triplet coefficient is reduced to prevent overly strong hard-negative constraints from interfering with reasoning sequence learning. Since the output distribution has stabilized after the cold-start stage and we no longer rely on reference policy constraints, Stage 2 removes the KL term ($\lambda_{\text{kl}}=0$). The total loss is:
\[
\mathcal{L}_{\text{stage2}}
=
2.4\,\mathcal{L}_{\text{sft}}
+ 1.0\,\mathcal{L}_{\text{nce}}
+ 6.9\,\mathcal{L}_{\text{tri}}
\]

\paragraph{Stage 3: Reasoning Optimization (GRPO)}
Stage 3 employs Group Relative Policy Optimization (GRPO). We sample multiple reasoning paths for the same query and optimize local logical quality through trial-and-error reinforcement learning with a reward system. The total reward comprises a ranking reward $R_{\text{rank}}$ and format constraints (gating/penalty), defined as:
\[
R_{\text{total}} = R_{\text{rank}} + R_{\text{format}}
\]
$R_{\text{rank}}$ adopts differentiable ranking to continualize discrete ranks, with the following formulation:
\[
\mathrm{Rank}(p) = 1 + \sum_{n\in N} \sigma\!\left(\frac{s(q,n)-s(q,p)}{\tau}\right)
\]
where $\sigma(\cdot)$ is the sigmoid function and $\tau$ is the temperature coefficient. Each sigmoid term above can be interpreted as a soft indicator function (varying continuously from 0 to 1) for ``whether the negative example surpasses the given positive example,'' making $\mathrm{Rank}(p)$ continuously differentiable. Based on this, we construct a normalized ranking reward:
\[
R_{\text{rank}}
=
1 - \frac{\mathbb{E}_{p\in P}\!\left[\log \mathrm{Rank}(p)\right]}{\log(|N|+1)}
\]
which falls within $[0,1]$ and provides a smoother, denser ranking quality signal. $R_{\text{format}}$ constrains the output to satisfy the ``reasoning sequence $\rightarrow$ \texttt{<emb\_token>}'' format and suppresses anomalous generation. GRPO updates parameters through relative reward advantages across sampled branches, thereby further reinforcing high-quality reasoning paths and stabilizing embedding representations during training.

\section{Experiments}

In this section, we conduct experiments to address the following research questions: How does the T1 model compare with existing text embedding models on reasoning-intensive retrieval tasks? What performance improvements does the T1 model achieve across its training stages? How does the T1 training strategy compare with contrastive learning?

\subsection{Experimental Setup}

We use Qwen3-Instruct-4B \cite{yang2025qwen3} as the base model for training the dense retrieval model. The training data for Stage 1 consists of 400K samples from the MS MARCO dataset\footnote{\url{https://huggingface.co/datasets/Tevatron/msmarco-passage}}, while Stages 2 and 3 use the 82K dataset provided by ReasonEmbed\footnote{\url{https://huggingface.co/datasets/hanhainebula/reason-embed-data}}. In Stage 2, we use GLM-4.5\footnote{\url{https://huggingface.co/zai-org/GLM-4.5}} to regenerate query reasoning. All training is conducted on 4 A800 GPUs, with Stage 1 requiring 20 hours, Stage 2 requiring 8 hours, and Stage 3 requiring 20 hours.

We use the BRIGHT evaluation benchmark as our primary evaluation dataset. This benchmark covers multiple reasoning-intensive domains including mathematics, science, and law. We adopt nDCG@10 as our primary metric, consistent with the BRIGHT standard.

For baseline models, we primarily select works that are representative in reasoning-intensive retrieval or the BRIGHT benchmark for comparison with T1: ReasonIR-8B \cite{shao2025reasonir}, which trains retrievers specifically for reasoning tasks; RaDeR-7B \cite{das2025rader}, based on a retrieval-augmented reasoning framework; and Seed1.5-Embedding\footnote{\url{https://seed1-5-embedding.github.io/}}, the embedding model from ByteDance. These models are compared with T1-4B under the same evaluation settings (original queries, nDCG@10) across BRIGHT subtasks.

\subsection{Main Results}

On the BRIGHT benchmark, T1-4B as a single model using only original queries achieves overall performance comparable to ``retrieval model + query rewriting or reranking'' combination pipelines, and significantly outperforms contrastive baselines and commercial embedding models. In other words, injecting reasoning capabilities through reinforcement learning during training enables retrieval effectiveness comparable to complex pipelines without the additional overhead of query rewriting or reranking at inference time. The results in Table~\ref{tab:bright-results} support this conclusion: under the ``Evaluate Retriever with Original Query'' setting, T1-4B achieves an average nDCG@10 of 37.1, surpassing ReasonIR-8B (24.4), RaDeR-7B (25.5), and Seed1.5-Embedding (27.2) by 12.7, 11.6, and 9.9 percentage points respectively. After applying GPT-4 REASON-query rewriting, ReasonIR-8B and RaDeR-7B only reach 29.9 and 29.2, still below T1-4B as a standalone model. Under the ``Evaluate Retriever with QwenRerank'' setting, ReasonIR with QwenRerank achieves 36.9, essentially on par with T1-4B (37.1), indicating that the single model without additional reranking modules achieves overall performance comparable to the retrieval--reranking pipeline.

\begin{table}[htbp]
  \centering
  \caption{Performance comparison of different models on BRIGHT}
  \label{tab:bright-results}
  \resizebox{0.95\textwidth}{!}{%
    \begin{tabular}{lcccccccccccc|c}
      \toprule
      \multirow{2}{*}{\textbf{Models}}
      & \multicolumn{7}{c}{\textbf{StackExchange}}
      & \textbf{Code}
      & \multicolumn{4}{c}{\textbf{Theorem-based}}
      & \multirow{2}{*}{\textbf{Avg}} \\
      \cmidrule(lr){2-8} \cmidrule(lr){9-9} \cmidrule(lr){10-13}
      & Bio & Earth & Econ & Psy & Rob & Stack & Sus
      & Pony
      & Leet & AoPS & TheoQ & TheoT
      & \\
      \midrule
      \multicolumn{14}{c}{\textit{Evaluate Retriever with Original Query}} \\
      \midrule
      ReasonIR-8B
      & 26.2 & 31.4 & 23.3 & 30.0 & 18.0 & 23.9 & 20.5
      & 10.5
      & 35.0 & \textbf{14.7} & 31.9 & 27.2
      & 24.4 \\
      RaDeR-7B
      & 34.6 & 38.9 & 22.1 & 33.0 & 14.8 & 22.5 & 23.7
      & 5.0
      & 37.3 & 10.2 & 28.4 & 35.1
      & 25.5 \\
      Seed1.5-Embedding
      & 34.8 & 46.9 & 23.4 & 31.6 & 19.1 & 25.4 & 21.0
      & 4.9
      & \textbf{43.2} & 12.2 & 33.3 & 30.5
      & 27.2 \\
      T1-4B
      & 57.4 & 54.8 & 30.6 & 48.2 & \textbf{33.1} & 36.4 & 35.6
      & 31.9
      & 14.9 & 11.9 & \textbf{41.6} & 48.5
      & 37.1 \\
      \midrule
      \multicolumn{14}{c}{\textit{Evaluate Retriever with GPT-4 REASON-query}} \\
      \midrule
      ReasonIR-8B
      & 43.6 & 42.9 & 32.7 & 38.8 & 20.9 & 25.8 & 27.5
      & 19.6
      & 31.5 & 7.4 & 33.1 & 35.7
      & 29.9 \\
      RaDeR-7B
      & 36.1 & 42.9 & 25.2 & 37.9 & 16.6 & 27.4 & 25.0
      & 11.9
      & 34.8 & 12.0 & 37.7 & 43.4
      & 29.2 \\
      \midrule
      \multicolumn{14}{c}{\textit{Evaluate Retriever with QwenRerank \cite{zhang2025qwen3}}} \\
      \midrule
      ReasonIR-8B
      & \textbf{58.2} & 53.2 & 32.0 & 43.6 & 28.8 & 37.6 & 36.0
      & \textbf{34.8}
      & 33.2 & 7.9 & 32.6 & 45.0
      & 36.9 \\
      RaDeR-7B
      & 58.0 & \textbf{59.2} & \textbf{33.0} & \textbf{49.4} & 31.8 & \textbf{39.0} & \textbf{36.4}
      & 33.3
      & 33.5 & 10.8 & 34.2 & \textbf{51.6}
      & \textbf{39.2} \\
      \bottomrule
    \end{tabular}%
  }
\end{table}

Across the seven StackExchange domains, the T1-4B single model is broadly on par with the top-performing combination of RaDeR-7B + QwenRerank, and achieves the table-wide highest score on Rob (33.1). T1-4B also performs notably well on Pony, TheoQ, and TheoT, though room for improvement remains on Leet and AoPS. These results are consistent with T1's architectural design: when encoding a query, T1-4B first generates a reasoning sequence before producing the embedding, effectively fusing ``query expansion'' and ``retrieval representation'' into a single end-to-end step. This explicit chain-of-thought is itself a form of reasoning, enabling the model to learn the semantic and logical path from the original query to the target document during training, thereby achieving overall performance comparable to multi-stage pipelines without relying on external query rewriting or reranking modules. Meanwhile, on tasks such as Leet and AoPS that depend heavily on domain-specific symbols and problem-solving patterns, the single model has not yet fully realized its advantage; future work may further unlock potential through domain-specific data or more fine-grained reward design.

\subsection{Analysis of the Results}

The three-stage training yields clear improvements in T1-4B's reasoning-intensive retrieval capabilities, with Stage~3 achieving the best performance on the vast majority of subtasks. The staircase-like improvement in average scores and Stage~3's leading performance on most subtasks confirm that task awareness establishes the retrieval format and foundational representations, reasoning data provides the main gains, and GRPO further calibrates ranking preferences. The data in Table~\ref{tab:t1-bright-results} support this conclusion: after establishing task awareness on MS~MARCO, Stage~1 achieves an average nDCG@10 of 23.0; Stage~2 rises to 35.7 (+12.7) after introducing ReasonEmbed data; and Stage~3 reaches 37.1 (+1.4) after GRPO. Except for Pony (36.6$\rightarrow$31.9) and AoPS (where Stage~1 achieves the highest at 12.1), Stage~3 attains the top score on all remaining 11 subtasks.

Further examination reveals that GRPO's benefits are unevenly distributed across subtasks. On subtasks that rely more heavily on multi-hop derivation and evidence chaining, such as Bio, Psy, and TheoT, the gains from Stage~3 over Stage~2 are more pronounced (e.g., Bio 53.8$\rightarrow$57.4, Psy 44.5$\rightarrow$48.2, TheoT 45.1$\rightarrow$48.5). In contrast, on tasks that are more oriented toward ``exact matching / symbol and format sensitivity,'' gains are smaller or even regressive (e.g., Code (Pony) 36.6$\rightarrow$31.9) or limited (e.g., AoPS 11.1$\rightarrow$11.9, still below Stage~1's 12.1). A plausible explanation is that the Stage~3 reward is primarily composed of differentiable ranking signals and format constraints, which tend to optimize ``reasoning-chain-driven relative ranking advantages'' but provide insufficient coverage of fine-grained matching preferences for code/symbol-heavy domains. Moreover, reinforcement learning is more susceptible to reward estimation noise in out-of-distribution domains, potentially causing over-optimization or mild forgetting. Therefore, such subtasks may require more targeted reward terms or domain-specific data to achieve stable improvements.

\begin{table}[htbp]
  \centering
  \caption{Performance comparison of T1-4B across training stages on BRIGHT}
  \label{tab:t1-bright-results}
  \resizebox{0.95\textwidth}{!}{%
    \begin{tabular}{lcccccccccccc|c}
      \toprule
      \multirow{2}{*}{\textbf{Models}}
      & \multicolumn{7}{c}{\textbf{StackExchange}}
      & \textbf{Code}
      & \multicolumn{4}{c}{\textbf{Theorem-based}}
      & \multirow{2}{*}{\textbf{Avg}} \\
      \cmidrule(lr){2-8} \cmidrule(lr){9-9} \cmidrule(lr){10-13}
      & Bio & Earth & Econ & Psy & Rob & Stack & Sus
      & Pony
      & Leet & AoPS & TheoQ & TheoT
      & \\
      \midrule
      T1-4B (Stage 1)
      & 23.8 & 39.2 & 18.4 & 30.0 & 21.3 & 23.5 & 19.8
      & 33.2
      & 6.7 & \textbf{12.1} & 27.5 & 20.5
      & 23.0 \\
      T1-4B (Stage 2)
      & 53.8 & 53.6 & 29.5 & 44.5 & 31.8 & 34.5 & 34.8
      & \textbf{36.6}
      & 12.7 & 11.1 & 40.7 & 45.1
      & 35.7 \\
      T1-4B (Stage 3)
      & \textbf{57.4} & \textbf{54.8} & \textbf{30.6} & \textbf{48.2} & \textbf{33.1} & \textbf{36.4} & \textbf{35.6}
      & 31.9
      & \textbf{14.9} & 11.9 & \textbf{41.6} & \textbf{48.5}
      & \textbf{37.1} \\
      \bottomrule
    \end{tabular}%
  }
\end{table}

\subsubsection{Comparison with Contrastive Learning}

Under the same 4B base model, T1 achieves a higher average nDCG@10 on BRIGHT than a retrieval model trained with contrastive learning, suggesting that training objectives based solely on representation similarity alignment may be insufficient for reasoning-intensive retrieval tasks. We attribute this gap to T1's incorporation of reasoning sequence generation and reinforcement learning rewards during training, which steers the model to attend more to implicit reasoning relationships when making relevance judgments. Table~\ref{tab:bright-results-comparison} shows that Contrastive Learning (4B) achieves an average nDCG@10 of 33.2, only marginally surpassing T1 on Econ (Contrastive Learning 31.2 vs.\ T1 30.6), while T1-4B (Stage 3) achieves an average nDCG@10 of 37.1 and exhibits more pronounced advantages on Bio, Earth, Psy, TheoQ, and TheoT.

\begin{table}[htbp]
  \centering
  \caption{Performance comparison of contrastive learning and T1 on BRIGHT}
  \label{tab:bright-results-comparison}
  \resizebox{0.95\textwidth}{!}{%
    \begin{tabular}{lcccccccccccc|c}
      \toprule
      \multirow{2}{*}{\textbf{Models}}
      & \multicolumn{7}{c}{\textbf{StackExchange}}
      & \textbf{Code}
      & \multicolumn{4}{c}{\textbf{Theorem-based}}
      & \multirow{2}{*}{\textbf{Avg}} \\
      \cmidrule(lr){2-8} \cmidrule(lr){9-9} \cmidrule(lr){10-13}
      & Bio & Earth & Econ & Psy & Rob & Stack & Sus
      & Pony
      & Leet & AoPS & TheoQ & TheoT
      & \\
      \midrule
      Contrastive Learning (4B)
      & 48.1 & 51.3 & \textbf{31.2} & 44.3 & 27.1 & 30.9 & 34.5
      & 24.3
      & 14.5 & 8.7 & 37.6 & 46.4
      & 33.2 \\
      T1-4B
      & \textbf{57.4} & \textbf{54.8} & 30.6 & \textbf{48.2} & \textbf{33.1} & \textbf{36.4} & \textbf{35.6}
      & \textbf{31.9}
      & \textbf{14.9} & \textbf{11.9} & \textbf{41.6} & \textbf{48.5}
      & \textbf{37.1} \\
      \bottomrule
    \end{tabular}%
  }
\end{table}

\section{Conclusion}

This paper presents T1, a reasoning-intensive retrieval model whose core contribution is shifting relevance modeling in retrieval from static representation consolidation via contrastive learning to dynamic reasoning generation. T1 endows the model with the ability to dynamically generate reasoning trajectories for each query, bridging implicit reasoning relationships through limited-step reasoning sequences on the query side and aggregating reasoning and semantic information into retrievable vector representations via \texttt{<emb\_token>}. Combined with three-stage curriculum training and GRPO's trial-and-error reward optimization, the model learns to dynamically select optimal derivation strategies for different queries. On the BRIGHT benchmark, T1-4B under the same 4B base setting outperforms embedding models trained with contrastive learning overall, achieves leading performance on multiple reasoning-intensive subtasks, and matches multi-stage pipeline performance without relying on additional query rewriting or reranking modules.

We believe reasoning-intensive retrieval has two complementary directions for future advancement. The first is to further strengthen effective trial-and-error through reinforcement learning with verifiable rewards: through more fine-grained, verifiable reward decomposition and more robust credit assignment mechanisms, training signals can be more directly aligned with ``the quality of derivation from query to evidence,'' while mitigating reward noise and over-optimization risks in out-of-distribution domains. The second is to explore \textit{test-time scaling}: trading controllable additional computation at inference time for greater reasoning depth (e.g., multi-branch reasoning, retrieval invocation decisions, or self-verification), and synergizing with training-time reasoning internalization to achieve more flexible performance--efficiency trade-offs under varying latency/cost budgets.

\bibliographystyle{plainnat}
\bibliography{references}

\appendix
\section{Appendix}

\subsection{Query-side Instruction Design}

On the query side, we adopt a designated query-side prompt and apply the Chat template. The actual input and expected reference output for Stage 1 are as follows:

\begin{promptblock}
\begin{lstlisting}
<|im_start|>system
You are an intelligent retrieval expert. Your goal is to generate the optimal vector representation for the user's query.<|im_end|>
<|im_start|>user
Instruct: Given a query, retrieve relevant passages that answer the query.\nquery: where is whitemarsh island<|im_end|>
<|im_start|>assistant
The embedding is <emb_token><|im_end|>
\end{lstlisting}
\end{promptblock}

We require the assistant to output the fixed suffix\allowbreak\ \texttt{The embedding is <emb\_token><|im\_end|>} as the reference output format after generating the reasoning sequence; the final query embedding is taken from the hidden state at the \texttt{<emb\_token>} position.

The instruction for Stage 2 differs from that of Stage 1. Stage 1 focuses on cold-start initialization and format consolidation, whereas Stage 2 enters the reasoning alignment phase, which requires explicitly instructing the model to generate a more structured reasoning sequence before outputting \texttt{<emb\_token>}, thereby improving reasoning alignment effectiveness and relevance modeling capability. Accordingly, the actual input and expected reference output for Stage 2 are as follows:

\begin{promptblock}
\begin{lstlisting}
QUERY_INSTRUCTION = """You are an intelligent retrieval expert. Your task is to enrich user input by increasing semantic depth in order to achieve more effective embedded representations. For each user input, please consider the following steps step by step:
1.Identify the core concepts and their interrelationships.
2.Incorporate key definitions and terms and expand necessary context-related synonyms.
3.Infer the key contents of the ideal target document.
After the analyzed content, you MUST end every response with <emb_token>."""
\end{lstlisting}
\end{promptblock}

The method for obtaining the final query embedding is the same as in Stage 1.

\subsection{Doc-side Instruction Design}

On the document side, we do not use \texttt{chat\_template}. The document-side instruction is concatenated with the document content, \texttt{<emb\_token>} is appended at the end, and the result is fed into the model for a single encoding-style forward pass. The hidden state at the \texttt{<emb\_token>} position is extracted as the document embedding.

\begin{promptblock}
\begin{lstlisting}
DOC_INSTRUCTION = """You are an intelligent retrieval expert. Your task is to analyze the input text and generate a comprehensive semantic vector embedding.
You should capture core concepts, factual details, and underlying logic to ensure the representation is robust for both keyword matching and complex reasoning tasks.
The embedding must represent the text's meaning accurately for high-quality retrieval."""
\end{lstlisting}
\end{promptblock}

\subsection{GLM-4.5 Prompt for Reasoning Trajectory Generation}

In Stage 2, we use GLM-4.5 to regenerate the reasoning trajectories accompanying the synthetic dataset from ReasonEmbed \cite{chen2025reasonembed}. The original reasoning trajectories are generally too long to serve effectively as alignment signals for ``limited-step reasoning sequences.'' To address this, we redesign the prompt by explicitly incorporating length control and quality constraints into the generation process, yielding shorter and more consistent reasoning trajectory data. The corresponding prompt is as follows:

\begin{promptblock}
  \begin{lstlisting}
QUERY_PROMPT = """
# Role
You are the world's most advanced search engine simulator. Your goal is to predict the **exact content**, **format**, and **style** of the ideal document that answers the user's query.

# Task
Based on the user's query, generate a **Hypothetical Document Passage** (approx. 100-200 words). Do not explain what the document *should* contain; instead, **write the document content directly**.

# Dynamic Style Guidelines (Crucial)
Analyze the query to determine the domain and adopt the matching style:

1.  **Coding & Technical Config** (e.g., Python, ROS, Pandas, Algorithms):
    * **Directly write code snippets**, CLI commands, directory trees, or log outputs.
    * Use specific library names, function names, and variable conventions (e.g., `self`, `df.interpolate`, `/catkin_ws`).
    * Do NOT provide beginner tutorials; provide the **solution code**.

2.  **Math, Logic & Physics** (e.g., Speed problems, Set theory):
    * **Solve the problem step-by-step**.
    * Use **LaTeX formatting** for formulas (e.g., $\mathcal{C}$, $\int$).
    * Show calculations, derivations, and proofs explicitly.

3.  **Academic, History & Science** (e.g., Oceanography, Banking Regulations, Sociology):
    * Write in a **dense, academic style**.
    * Hallucinate/Predict specific **dates, acts, legislation, citations, and technical terminology** (e.g., "DIDMCA", "halocline", "structural barriers").
    * Mimic the tone of a research paper abstract or a textbook excerpt.

4.  **General/Hobbyist** (e.g., Aquaponics):
    * Write in an informative blog post or forum answer style.
    * Focus on **mechanisms** and **practical functionality**.

# Constraints
* **NO** introductory filler (e.g., "Here is the code...", "The document discusses...").
* **NO** dictionary definitions unless explicitly asked.
* **Start directly** with the content.

# Input Query:
"""
\end{lstlisting}
\end{promptblock}

\end{document}